\documentclass[a4paper,10pt]{cpc-hepnp}
\usepackage{CJK,upgreek,fancyhdr}
\usepackage{subfigure}
\usepackage{multicol}
\usepackage{booktabs}
\usepackage{amssymb,bm,mathrsfs,bbm,amscd}
\usepackage[tbtags]{amsmath}
\usepackage{lastpage}
\usepackage{graphicx}
\usepackage{multirow}

\DeclareGraphicsExtensions{.pdf,.jpeg,.png}

\usepackage{epstopdf}
\usepackage[english]{babel}
\usepackage[colorlinks=false,pdfpagemode=FullScreen,,setpagesize=off,pdfborder={0 0 0}]{hyperref}
\usepackage{overpic}
\usepackage{lineno}
\usepackage{color}

\newcommand{\ee}{e^+e^-}

\newcommand{\mev}{\,\mathrm{MeV}}

\newcommand{\mevcc}{\,\mathrm{MeV}/c^2}
\newcommand{\gev}{\,\mathrm{GeV}}

\newcommand{\gevcc}{\,\mathrm{GeV}/c^2}

\newcommand {\ie}       {\emph{i}.\emph{e}.}

\newcommand{\zcsm}{Z_{cs}(3985)^{-}}
\newcommand{\zcs}{Z_{cs}(3985)}
\newcommand{\zcspm}{Z_{cs}^{\prime -}}

\newcommand{\zcsprime}{Z_{cs}^{\prime}}

\newcommand{\DDstar}{D_s^{*-} D^{*0}}
\newcommand{\Dsm}{D_{s}^{-}}
\newcommand{\Dsstm}{D_{s}^{*-}}
\newcommand{\Dzero}{D^{0}}
\newcommand{\Dzst}{D^{*0}}

\newcommand{\DDhs}{D_{s}^{*-}D_{s1}(2536)^{+}}

\newcommand{\kp}{K^{+}}
\newcommand{\km}{K^{-}}
\newcommand{\kdsstdstzero}{K^{+}D_{s}^{*-}D^{*0}}

\begin{document}
\begin{CJK*}{UTF8}{gkai}

\fancyhead[c]{\small Chinese Physics C~~~Vol. XX, No. X (2022) XXXXXX}
\fancyfoot[C]{\small 010201-\thepage}

\footnotetext[0]{Received xxxx June xxxx}

\title{Search for hidden-charm tetraquark with strangeness in $e^{+}e^{-}\rightarrow K^+ D_{s}^{*-} D^{*0}+c.c.$
\thanks{
Supported in part by National Key R\&D Program of China under Contracts Nos. 2020YFA0406400, 2020YFA0406300; National Natural Science Foundation of China (NSFC) under Contracts Nos. 11635010, 11735014, 11805086, 11835012, 11935015, 11935016, 11935018, 11961141012, 12022510, 12025502, 12035009, 12035013, 12192260, 12192261, 12192262, 12192263, 12192264, 12192265; the Chinese Academy of Sciences (CAS) Large-Scale Scientific Facility Program; Joint Large-Scale Scientific Facility Funds of the NSFC and CAS under Contract No. U1832207; the CAS Center for Excellence in Particle Physics (CCEPP); 100 Talents Program of CAS; Fundamental Research Funds for the Central Universities, Lanzhou University, University of Chinese Academy of Sciences; The Institute of Nuclear and Particle Physics (INPAC) and Shanghai Key Laboratory for Particle Physics and Cosmology; ERC under Contract No. 758462; European Union's Horizon 2020 research and innovation programme under Marie Sklodowska-Curie grant agreement under Contract No. 894790; German Research Foundation DFG under Contracts Nos. 443159800, Collaborative Research Center CRC 1044, GRK 2149; Istituto Nazionale di Fisica Nucleare, Italy; Ministry of Development of Turkey under Contract No. DPT2006K-120470; National Science and Technology fund; National Science Research and Innovation Fund (NSRF) via the Program Management Unit for Human Resources \& Institutional Development, Research and Innovation under Contract No. B16F640076; Olle Engkvist Foundation under Contract No. 200-0605; STFC (United Kingdom); Suranaree University of Technology (SUT), Thailand Science Research and Innovation (TSRI), and National Science Research and Innovation Fund (NSRF) under Contract No. 160355; The Royal Society, UK under Contracts Nos. DH140054, DH160214; The Swedish Research Council; U. S. Department of Energy under Contract No. DE-FG02-05ER41374.
}
}

\maketitle
\begin{center} 
\begin{small}
\begin{center}
M.~Ablikim(麦迪娜)$^{1}$, M.~N.~Achasov$^{12,b}$, P.~Adlarson$^{72}$, M.~Albrecht$^{4}$, R.~Aliberti$^{33}$, A.~Amoroso$^{71A,71C}$, M.~R.~An(安美儒)$^{37}$, Q.~An(安琪)$^{68,55}$, Y.~Bai(白羽)$^{54}$, O.~Bakina$^{34}$
% ...
\\
\vspace{0.2cm}
(BESIII Collaboration)\\
\vspace{0.2cm} {\it
$^{1}$ Institute of High Energy Physics, Beijing 100049, People's Republic of China\\
% ...
$^{4}$ Bochum Ruhr-University, D-44780 Bochum, Germany\\
% ...
$^{12}$ G.I. Budker Institute of Nuclear Physics SB RAS (BINP), Novosibirsk 630090, Russia\\
% ...
$^{33}$ Johannes Gutenberg University of Mainz, Johann-Joachim-Becher-Weg 45, D-55099 Mainz, Germany\\
$^{34}$ Joint Institute for Nuclear Research, 141980 Dubna, Moscow region, Russia\\
% ...
$^{37}$ Liaoning Normal University, Dalian 116029, People's Republic of China\\
% ...
$^{54}$ Southeast University, Nanjing 211100, People's Republic of China\\
$^{55}$ State Key Laboratory of Particle Detection and Electronics, Beijing 100049, Hefei 230026, People's Republic of China\\
% ...
$^{68}$ University of Science and Technology of China, Hefei 230026, People's Republic of China\\
% ...
$^{71}$ University of Turin and INFN, (A)University of Turin, I-10125, Turin, Italy; (B)University of Eastern Piedmont, I-15121, Alessandria, Italy; (C)INFN, I-10125, Turin, Italy\\
$^{72}$ Uppsala University, Box 516, SE-75120 Uppsala, Sweden\\
% ...
$^{b}$ Also at the Novosibirsk State University, Novosibirsk, 630090, Russia\\
% ...
}\end{center}

\vspace{0.4cm}
\end{small}
\end{center}


\begin{abstract}
We report a search for a heavier partner of the recently observed $Z_{cs}(3985)^{-}$ state, denoted as $Z_{cs}^{\prime -}$, in the process $e^{+} e^{-}\rightarrow K^{+}D_{s}^{*-}D^{* 0}+c.c.$, based on $e^+e^-$ collision data collected at the center-of-mass energies of $\sqrt{s}=4.661$, 4.682 and 4.699 GeV with the BESIII detector.
The $Z_{cs}^{\prime -}$ is of interest as it is expected to be a candidate for a hidden-charm and open-strange tetraquark.
A partial-reconstruction technique is used to isolate $K^+$ recoil-mass spectra, which are probed for a potential contribution from $Z_{cs}^{\prime -}\to D_{s}^{*-}D^{* 0}$ ($c.c.$).
We find an excess of $Z_{cs}^{\prime -}\rightarrow D_{s}^{*-}D^{*0}$ ($c.c.$) candidates with a significance of $2.1\sigma$, after considering systematic uncertainties, at a mass of $(4123.5\pm0.7_\mathrm{stat.}\pm4.7_\mathrm{syst.}) \mevcc$.
As the data set is limited in size, the upper limits are evaluated at the 90\% confidence level on the product of the Born cross sections ($\sigma^{\mathrm{Born}}$) and the branching fraction ($\mathcal{B}$) of $Z_{cs}^{\prime-}\rightarrow D_{s}^{*-}D^{* 0}$, under different assumptions of the $Z_{cs}^{\prime -}$ mass from 4.120 to 4.140 MeV and of the width from 10 to 50 MeV at the three center-of-mass energies. 
The upper limits of $\sigma^{\rm Born}\cdot\mathcal{B}$ are found to be at the level of $\mathcal{O}(1)$ pb at each energy.
Larger data samples are needed to confirm the  $Z_{cs}^{\prime -}$ state and clarify its nature in the coming years.
\end{abstract}



\newpage
\begin{multicols}{2}

\section{Introduction} 
The many examples of exotic-hadron candidates that have been reported in recent years provide important constraints on Quantum Chromodynamic (QCD) phenomenological models and further deepen our understanding of the strong interaction, offering insights beyond those available through studies of conventional mesons and baryons~\cite{chen_updated_2022,XYZ_review1,XYZ_review2, XYZ_review3, XYZ_review4}. 
In 2021, the first observation of a charged hidden-charm tetraquark candidate with strangeness, named $Z_{cs}(3985)^{-}$, decaying into $D_s^{-}D^{*0}$ and $D_s^{*-}D^0$, was reported in the process $\ee\to K^+ (D_s^{-} D^{*0} + D_s^{*-} D^{0})$ by the BESIII Collaboration~\cite{BESIII:Zcs}.
Here and elsewhere, the charge-conjugation mode is always implied, unless otherwise stated.
The mass and width of the $Z_{cs}(3985)^-$ were measured to be $(3982.5_{-2.6}^{+1.8} \pm 2.1)\mevcc$ and $(12.8_{-4.4}^{+5.3}\pm3.0)\mev$, respectively.
The LHCb Collaboration later reported two more charged $Z_{cs}(4000)^-$ and $Z_{c s}(4220)^-$ states in the mass spectra of $J/\psi K^{-}$ in $B$ decays~\cite{LHCb:Zcs}.
These observations have stimulated much theoretical discussion, especially, on the relationship between the $Z_{cs}(3985)^{-}$, $Z_{cs}(4000)^{-}$ and $Z_{cs}(4220)^{-}$ states~\cite{Karliner:2021enr,chen_updated_2022}. One notable observation is that the width of the $Z_{cs}(3985)^{-}$ is much narrower than those of the $Z_{cs}(4000)^{-}$ and $Z_{cs}(4220)^{-}$. This indicates that the $Z_{cs}(3985)^{-}$ and $Z_{cs}(4000)^{-}$ may be not the same state, although they are close-lying in mass~\cite{Maiani:2021tri,Meng:2021rdg}.
Earlier this year, the BESIII Collaboration reported ~\cite{BESIII:2022qzr} evidence for the neutral $Z_{cs}(3985)^0$ state in the process $\ee\to K_{S}^{0}(D_s^+ D^{*-}+D_s^{*+} D^{-})$, whose mass and width are close to those of the charged $Z_{cs}(3985)^{-}$. Furthermore, the relative Born cross sections for the production of these two states agree with the expectations of isospin symmetry. 
Hence, the $Z_{cs}(3985)^0$ is regarded as the isospin partner of the $Z_{cs}(3985)^{-}$~\cite{BESIII:2022qzr}.

As discussed in Refs.~\cite{ZcsH1, ZcsH2, ZcsH3, ZcsH4, ZcsH5, ZcsH6, ZcsH_Mass1, ZcsH_Mass2, ZcsH_Mass3, ZcsH_Mass4, ZcsH_Mass5, ZcsH_Mass6, ZcsH_Mass7, ZcsH_Mass8, ZcsH_Mass9, ZcsH_Mass10, ZcsH_Mass11, ZcsH_Mass12, ZcsH_Mass13, ZcsH_Mass14, ZcsH_Mass15, Wang:2022fdu}, heavier partners of the $Z_{cs}(3985)^-$ might exist, which are denoted as $Z_{cs}'^-$.
The $Z_{cs}'^-$ state is expected to decay with large rate to the final state of $D_s^{*-}D^{*0}$, and has masses ranging from 4120 to 4200 MeV/$c^2$ in different models~\cite{ZcsH_Mass1, ZcsH_Mass2, ZcsH_Mass3, ZcsH_Mass4, ZcsH_Mass5, ZcsH_Mass6, ZcsH_Mass7, ZcsH_Mass8, ZcsH_Mass9, ZcsH_Mass10, ZcsH_Mass11, ZcsH_Mass12, ZcsH_Mass13, ZcsH_Mass14, ZcsH_Mass15}. For example, $\zcspm$ is assumed to be a $\DDstar$ molecule with a mass of 4140$\mevcc$ in Ref.~\cite{ZcsH_Mass1} while the heavy $SU(3)_f$ partner of $Z_{cs}(3985)^-$ is predicted to have a mass of 4124$\mevcc$ in chiral effective field theory~\cite{ZcsH_Mass5}. Thus, it is crucial to search for the $Z_{cs}'^-$ state and measure its properties to distinguish between these models. 

In this paper, we search for $Z_{cs}'^-$ state in the process of $e^+e^-\rightarrow K^+\DDstar$, taking advantage of $e^+e^-$ collision data collected with the BESIII detector at the center-of-mass energies $\sqrt{s}=4.661$, 4.682, and 4.699 GeV~\cite{BESIII:2022ulv} . The corresponding integrated luminosity of these data samples is 2.7 fb$^{-1}$~\cite{BESIII:2022ulv}.

\section{BESIII detector}
The BESIII detector~\cite{BESIII:2009fln} records symmetric $e^+e^-$ collisions provided by the BEPCII storage ring~\cite{Yu:2016cof} in the center-of-mass energy range from 2.0 to 4.95~GeV, with a peak luminosity of $1 \times 10^{33}\;\text{cm}^{-2}\text{s}^{-1}$ achieved at $\sqrt{s}=3.77\;\text{GeV}$.
BESIII has collected many data samples in this energy region~\cite{BESIII:2020nme}. 
The cylindrical core of the BESIII detector covers 93\% of the full solid angle and consists of a helium-based multilayer drift chamber~(MDC), a plastic scintillator time-of-flight system~(TOF), and a CsI(Tl) electromagnetic calorimeter~(EMC), which are all enclosed in a superconducting solenoidal magnet providing a 1.0~T magnetic field. 
The solenoid is supported by an octagonal flux-return yoke with resistive plate counter muon identification modules interleaved with steel. 
The charged-particle momentum resolution at $1~{\rm GeV}/c$ is $0.5\%$, and the specific ionization energy loss~$\mathrm{d}E/\mathrm{d}x$ resolution is $6\%$ for electrons from Bhabha scattering. 
The EMC measures photon energies with a resolution of $2.5\%$ ($5\%$) at $1$~GeV in the barrel (end-cap) region. 
The time resolution of the TOF barrel part is 68~ps, while that of the end-cap part is 110~ps. 
The end-cap TOF system was upgraded in 2015 using multi-gap resistive plate chamber technology, providing a time resolution of 60~ps~\cite{li_study_2017,guo_study_2017,Cao:2020ibk}. 

\section{Monte Carlo simulation}
Simulated data samples produced with a {\sc geant4}-based~\cite{GEANT4:2002zbu} Monte Carlo (MC) package, 
which includes the geometric description of the BESIII detector~\cite{GDMLMethod,BesGDML,geometry} and the detector responses,
are used to determine detection efficiencies and to estimate backgrounds. 
The simulation models the beam-energy spread and initial-state radiation (ISR) in the $e^+e^-$ annihilations with the generator {\sc kkmc}~\cite{Jadach:2000ir,Jadach:1999vf}. 
For the three-body non-resonant (NR) signal process, $\ee\rightarrow\kdsstdstzero$, the final-state particles are simulated assuming non-resonant production~\cite{BESIII:2020nme}.
For the simulation of the $\zcspm$ signal process, $\ee\rightarrow\kp\zcspm\to\kdsstdstzero$, the spin-parity $J^{P}$ of the $\zcspm$ is assumed to be $1^{+}$ in accordance with Refs.~\cite{BESIII:Zcs, LHCb:Zcs}, as the corresponding production and subsequent decay process are both dominated by $S$ wave.
However, other spin-parity assignments are allowed, and considered in the assignment of systematic uncertainties.

\section{Event selection}
To ensure a high selection efficiency, we employ a partial-reconstruction method, in which we use two tag modes to identify the process of $\ee\rightarrow\kdsstdstzero$. 
In the $\Dsm$-tag mode, we reconstruct the prompt $\kp$ and the decay products of the $\Dsm$ via the $\Dsm\rightarrow \kp\km\pi^{+}$ or $K_{S}^{0}\km$ processes.
The strategy for the $\Dzst$-tag mode is similar to the $\Dsm$-tag case but here the $\Dzst$ is reconstructed via $\Dzst\to \Dzero\pi^{0}$, and $\Dzero\rightarrow \km\pi^{+}$ or $\km\pi^{+}\pi^{+}\pi^{-}$, $\pi^{0}\to\gamma\gamma$ process.

Each charged track is required to have a polar angle $\theta$ in the range $|\cos\theta|<0.93$.
For all such tracks, apart from those used to construct $K^0_S$ candidates, the distances of closest approach to the interaction point ($V_{z}$) are required to be less than 10\,cm along the beam direction and less than 1\,cm in the plane perpendicular to the beam ($V_{r}$).
The $\mathrm{d}E/\mathrm{d}x$ in the MDC and the time of flight information measured in the TOF are
used to calculate particle identification (PID) likelihood values for the pion ($\mathcal{L}_{\pi}$) and kaon ($\mathcal{L}_{K}$) hypotheses. 
Pion candidates are selected by requiring $\mathcal{L}_{\pi}>\mathcal{L}_{K}$, and kaon candidates are required to satisfy $\mathcal{L}_{K}>\mathcal{L}_{\pi}$.

Photon candidates are reconstructed from isolated clusters in the
EMC in the regions $|\!\cos\theta|\le0.80$ (barrel) or $0.86\le|\!\cos\theta|\le 0.92$ (end cap).
The deposited energy of a cluster is required to be greater than 25 (50) MeV in the barrel (end-cap) region.
To suppress electronic noise and energy deposits unrelated to the events, the difference between the EMC time and the event start time is required to be within (0, 700)~ns. The $\pi^{0}$ candidates are reconstructed from all combinations of photon pairs and the diphoton invariant mass, $M_{\gamma\gamma}$, is required to satisfy $0.115<M_{\gamma\gamma}<0.150\ \gevcc$.

Candidates for $K_{S}^{0}$ mesons are reconstructed from two oppositely charged tracks, with no PID requirement;
A looser requirement of $V_{z}<20$\,cm is applied on these tracks than $R_{z}<10$\,cm while no $V_{r}$ requirement is applied.
Furthermore, there is usually a detectable displacement before the decay of $K_{S}^{0}$ meson due to its relatively long lifetime. 
Therefore, the decay length and corresponding uncertainty of $K_{S}^{0}$ candidates are required to satisfy $L/\sigma_{L}>2$, which suppresses the prompt $\pi^{+}\pi^{-}$ combinatorial background~\cite{BESIII:2019ugy}. 
The $K_{S}^{0}$ meson candidates that lie within the invariant-mass window $0.485<M(\pi^{+}\pi^{-})<0.511\ \gevcc$ are retained.

\begin{figure*}[tph]
    \centering
    \includegraphics[width=\textwidth]{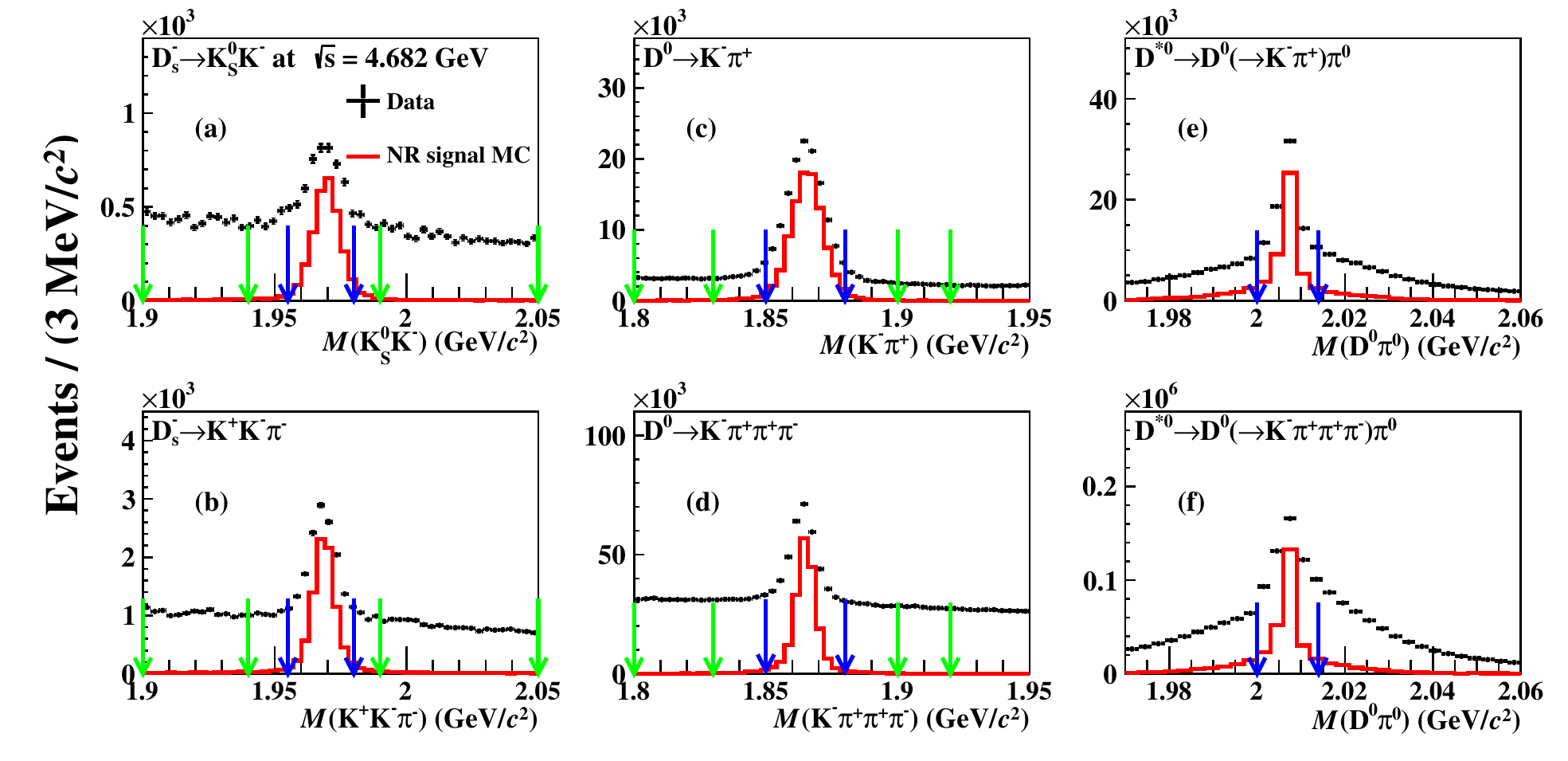}
    \caption{The (a)\,$\kp\km\pi^{-}$, (b)\,$K_{S}^0\km$, (c)\,$\km\pi^{+}$, (d)\,$\km\pi^+\pi^+\pi^-$ invariant-mass spectra and (e,f) the corresponding $\Dzero\pi^0$ mass spectra. The black dots are the data sample at $\sqrt{s}=4.682\ \gev$, the red histograms are the NR signal MC samples, and the blue (green) arrows indicate the signal (sideband) regions. }
    \label{fig:d_mass}
\end{figure*}

The $\Dsm$ and $\Dzst$ candidates for all decay modes are reconstructed by considering all combinations of selected charged track, $\pi^{0}$ and $K_{S}^{0}$ candidates. 
To improve the signal purity in the selected $D^-_s\to K^+K^-\pi^-$ candidates, we only retain the $\Dsm$ formed by $\Dsm\rightarrow \phi\pi^{-}$ or $\Dsm\rightarrow K^{*}(892)^{0}\km$ by requiring $M(\kp\km)<1.05\ \gevcc$ or $0.850<M(\kp\pi^-)<0.930\ \gevcc$.
Figure~\ref{fig:d_mass} shows the invariant mass spectra of the $\Dsm$ and $\Dzero$ candidates at $\sqrt{s}=4.682\ \gev$, in which the signal peaks are clearly evident.
To select the $D_s^-$ and $D^0$ signal candidates, their reconstructed invariant masses are restricted to lie in the regions $(1.955, 1.980)\ \gevcc$ and $(1.850, 1.880)\ \gevcc$, respectively.
Furthermore, in the selection of $\Dzst\rightarrow\Dzero\pi^{0}$ decays, the invariant mass of the $\Dzero$ and $\pi^0$ candidate, $M(\Dzero\pi^{0})$, is calculated with the reconstructed momenta of the final states and the known $m(\pi^{0})$ mass from the PDG~\cite{ParticleDataGroup:2020ssz} in order to improve mass resolution~\cite{BESIII:2013mhi}, as shown in Fig.~\ref{fig:d_mass}.
In case there are multiple $\Dzst$ candidates per event, only the candidate with invariant mass in the region $(2.000, 2.014)\ \gevcc$ and closest to the known $\Dzst$ mass is kept.

Figure~\ref{fig:RM KD selection} shows the spectrum of the $\kp\Dsm$ ($\kp\Dzst$) recoil mass, $RM(\kp\Dsm)$ ($RM(\kp\Dzst)$), for the $\Dsm$ ($\Dzst$) candidate events from the data samples at $\sqrt{s}=4.682\ \gev$.
Here, $RM(X)=\|p_{e^{+}e^{-}}-p_{X}\|$, where $p_{e^{+}e^{-}}$ is the four-momentum of the initial $e^{+}e^{-}$ system and $p_{X}$ is the four-momentum of the system $X$.
The $RM(\kp\Dsm)$ ($RM(\kp\Dzst)$) is corrected by substituting the reconstructed $\Dsm$ ($\Dzst$) mass with the known mass $m(\Dsm)$ ($m(\Dzst)$) from the PDG~\cite{ParticleDataGroup:2020ssz}.
In the $RM(\kp\Dsm)$ spectrum with the $\Dsm$-tag method shown in Fig.~\ref{fig:RM KD selection}, the peak around the mass of $D^{*0}$ of $2006.85\,\mevcc$~\cite{ParticleDataGroup:2020ssz} is due to background process $\ee\to K^+\Dsm\Dzst$. The process $\ee\to K^+\Dsstm\Dzero$ also contributes to this peak but shows a broader structure. These processes will not dilute the enhancement above $2.11\ \gevcc$ corresponding to the combination of the unreconstructed $\Dzst$ and $\gamma$~($\pi^{0}$) from the $\Dsstm$ decays in the signal process $\ee\rightarrow\kdsstdstzero$.
Based on this feature, we further improve the signal purity with the requirement $R(\kp\Dsm)>2.11\ \gevcc$.
In the $\Dzst$-tag method, a clear peak is seen in the $RM(\kp\Dzst)$ spectrum at the nominal $\Dsstm$ mass.
Therefore, we require $RM(\kp\Dzst)$ to be in the interval $(2.102, 2.122)\ \gevcc$ to isolate the $\Dsstm$ signals.
The double-counting rate between these two tag methods is negligible after applying all the above selection criteria. 
The resulting $RM(\kp)$ spectra for the surviving events at various energy points, as shown in Fig.~\ref{fig:RM_K fit}, are then probed for contributions from the $\zcspm$ state. Estimated background processes are also overlaid in Fig.~\ref{fig:RM_K fit}.


\begin{center}
\includegraphics[width=0.4\textwidth]{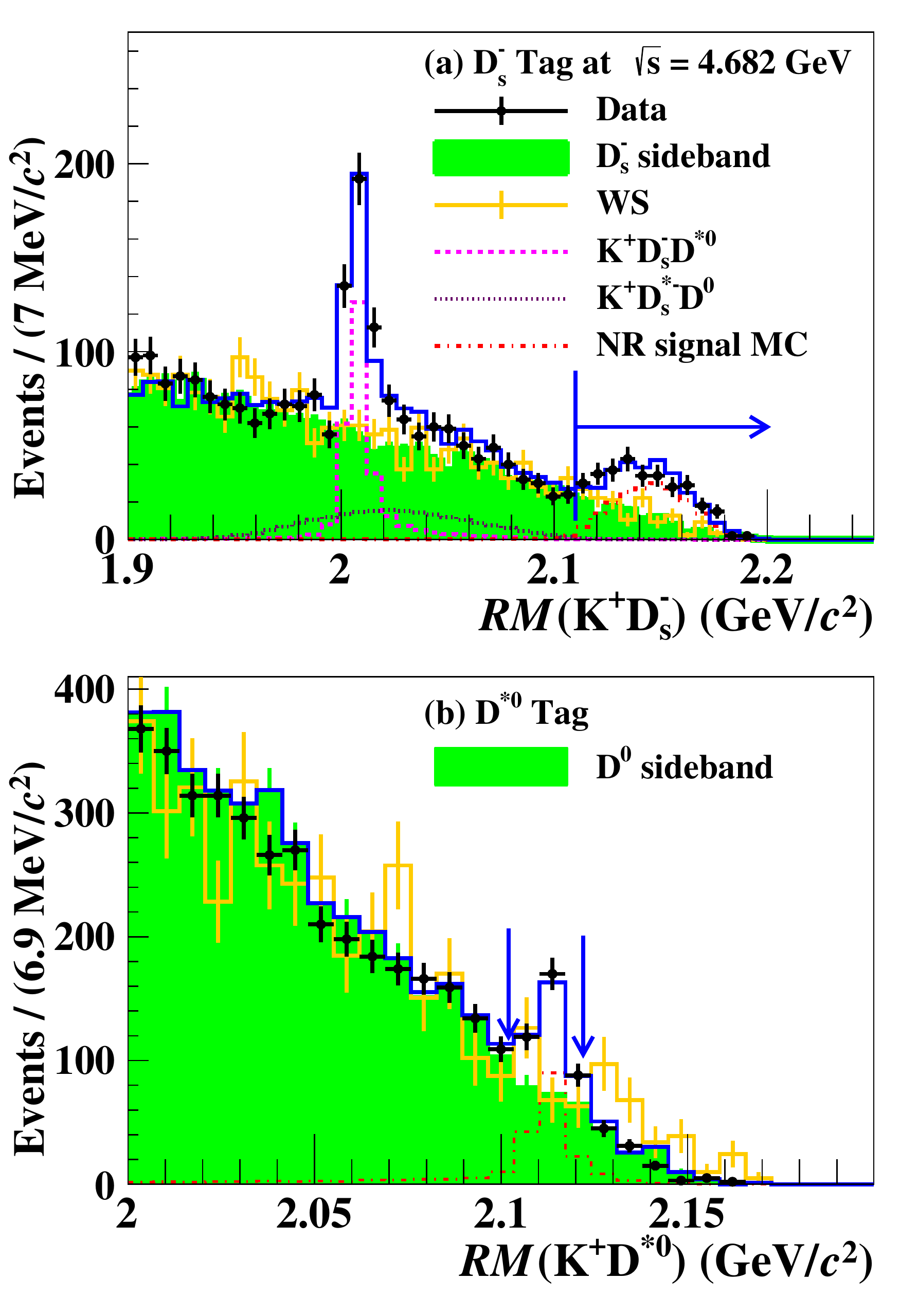}
\figcaption{The spectra of (a) recoil mass $RM(\kp\Dsm)$ of $\kp\Dsm$ and (b) recoil mass $RM(\kp\Dzst)$ of $\kp\Dzst$ for the $\Dsm(\Dzst)$ candidate events at $\sqrt{s}=4.682\ \gev$. The scaled background control samples of wrong sign (WS) events and $D_{(s)}$ sideband events are overlaid to represent the combinatorial backgrounds. The scaled NR signal MC is also overlaid. The blue lines are the sum of spectra of scaled $\Dsm$ sideband, $\ee\to K^+\Dsm\Dzst$ and $\ee\to K^+\Dsstm\Dzero$ and NR signal MC. The blue arrows indicate the selected region for the $\ee\to\kdsstdstzero$ candidates.}
\label{fig:RM KD selection}
\end{center}

The combinatorial background shape is derived from the kernel estimation~\cite{Cranmer:2000du} of a mixed background sample formed from wrong-sign (WS) combinations and $\Dsm(\Dzero)$ mass sideband events. 
The WS sample consists of $\km$ and $\Dsm$($\bar{D}^{*0}$) combinations, rather than the right-sign $\kp$ and $\Dsm$($\bar{D}^{*0}$) combinations for the correct signal selection.
The sideband samples for the $\Dsm$ and $\Dzero$ events are defined as $(1.90,1.94)\|(1.99,2.05)\ \gevcc$ in the $\Dsm$ invariant-mass spectrum, and $(1.80,1.83)\|(1.90,1.92)\ \gevcc$ in the $\Dzero$ invariant-mass spectrum, as depicted in Fig.~\ref{fig:d_mass}.
When forming the mixed background sample, we scale the contribution of the $\Dsm(\Dzero)$ mass sideband events to equal that of the WS combinations. 
The number of background events within the signal region is estimated by a fit to the $RM(\kp\Dsm)$ $(RM(\kp\Dzst))$ spectrum of the sideband events with the shape extracted from the mixed sample, as described in Ref.~\cite{BESIII:2022qzr}.
In the fit to the $RM(\kp\Dsm)$ spectrum, two additional contributions from the processes $\ee\rightarrow \kp\Dsm\Dzst$ and $\kp\Dsstm\Dzero$ are taken into account, whose shapes are derived from MC simulations.
Table~\ref{tab:bkg} lists the numbers of combinatorial background events in Fig.~\ref{fig:RM_K fit} for the two tag methods at various energy points.

\begin{table*}[tph]
    \caption{Estimated contributions from the combinatorial background and $\DDhs$ events in the final sample of $\ee\to\kp\Dsstm\Dzst$ candidates for the two tag methods at various energy points. The uncertainties are statistical.}
    \begin{center}\begin{tabular}{ccccc}
        \hline\hline
        Background source              &Tag         &$4.661\ \gev$ &$4.682\ \gev$ &$4.699\ \gev$ \\\hline
        \multirow{2}*{Comb. bkg} &$\Dsm$-tag  &$27\pm5$   &$120\pm11$ &$54\pm7$   \\
                                 &$\Dzst$-tag &$33\pm6$   &$216\pm15$ &103$\pm10$ \\\hline
        \multirow{2}*{$\DDhs$}   &$\Dsm$-tag  &$18\pm7$   &$117\pm27$ &$52\pm13$  \\
                                 &$\Dzst$-tag &$15\pm6$   &$91\pm21$  &$33\pm9$   \\
        \hline\hline
    \end{tabular}\end{center}
    \label{tab:bkg}
\end{table*}

Studies are performed with MC simulation to understand the potential contributions to the $RM(K^+)$ spectra from the excited charmed mesons $D_{(s)}^{**}$, $\ie$, $\ee\rightarrow \Dsstm D_s^{**+}(\rightarrow \Dzst\kp)$ or $\Dzst D^{**0}(\rightarrow \Dsstm\kp)$, which have the same final states as the signal processes. These studies account for all known $D_{(s)}^{**}$ states~\cite{ParticleDataGroup:2020ssz, LHCb:2019juy}.
It is found that the channel $\ee\to\DDhs(\rightarrow\Dzst\kp)$ gives a broad contribution at the higher mass region in the $RM(\kp)$ spectra, as shown in Fig.~\ref{fig:RM_K fit}.
We evaluate the size of this contribution according to the measured production cross sections~\cite{BESIII:2021xrz}, to yield the numbers listed in Table~\ref{tab:bkg}.
Other potential $D_{(s)}^{**}$ contributions behave in a similar way as the NR signal processes.

As shown in Fig.~\ref{fig:RM_K fit}, there is a non-trivial structure around $4.125\gevcc$ in the $RM(\kp)$ spectra when excluding the contributions from combinatorial backgrounds and $\DDhs$ components.
Both the $\Dsm$-tag and $\Dzst$-tag samples show an enhancement above the estimated backgrounds and the NR process, in particular at $\sqrt{s}=4.682\gev$.

\section{Measurement of resonance parameters}

\begin{figure*}
    \centering
    \includegraphics[width=0.8\textwidth]{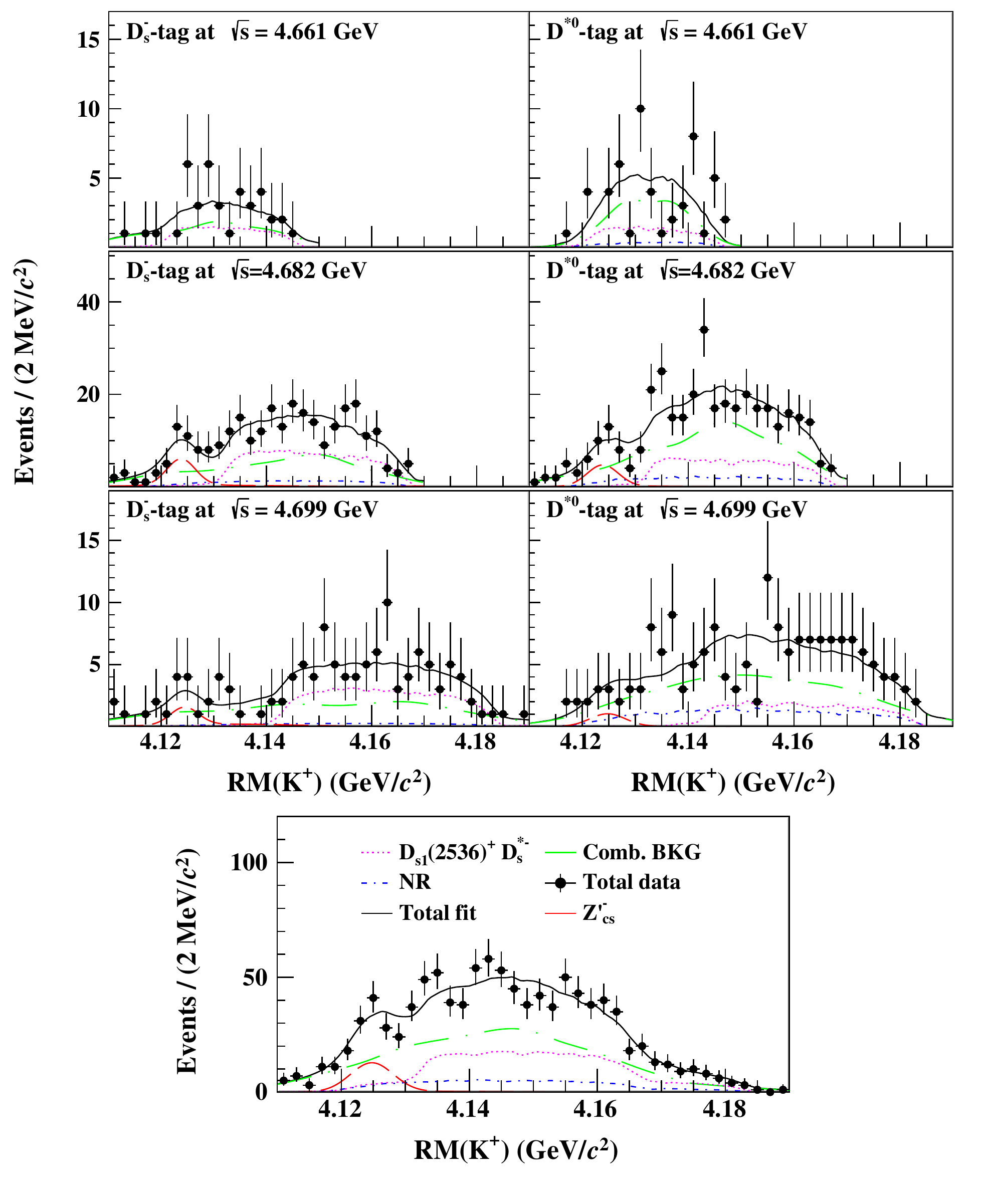}
    \caption{The spectra of $\kp$ recoil mass from $\Dzst$- and $\Dsm$-tag methods after all selection criteria at $\sqrt{s}=4.661,4.682$ and $4.699\ \gev$. The simultaneous fit results are overlaid, where the red dashed lines are $\zcspm$ signals, the blue dashed-dotted lines are NR processes, the green dashed-dotted lines are combinatorial background, the purple dashed lines represent the process $\ee\to\DDhs$ and the black solid lines are the sum of all components. }
    \label{fig:RM_K fit}
\end{figure*}

We consider the enhancement around $4.125\ \gevcc$ in the $RM(K^+)$ spectra to be a $\Dsstm\Dzst$ resonance, which we denote as $\zcspm$.
To estimate the number of $\zcspm$ events in the sample, we perform a simultaneous unbinned maximum-likelihood fit to the $RM(\kp)$ spectra for the three energy points with the two tag methods, as shown in Fig.~\ref{fig:RM_K fit}.
The $\zcspm$ component is modeled with an efficiency-weighted Breit-Wigner function convolved with a resolution function estimated from MC simulations. The Breit-Wigner function assumes an $S$-wave in the following form
\begin{equation*}
    \mathcal{F}(M) \propto \Big | \frac{ \sqrt{q \cdot p}}{M^2-m_{0}^2 + im_{0}\Gamma(M)} \Big |^2,\nonumber
\end{equation*}
with $\Gamma(M)=\Gamma_{0}\cdot\frac{p}{p^{*}}\cdot\frac{m_{0}}{M}$.
Here, $M$ is the reconstructed mass, $m_{0}$ is the resonance mass, 
$\Gamma_{0}$ is the width,
$q$ is the $\kp$ momentum in the initial $\ee$ system,
$p$ is the $\Dsstm$ momentum in the rest frame of the $\Dsstm\Dzst$ system, and
$p^{*}$ is the $\Dsstm$ momentum in the rest frame of the $\Dsstm\Dzst$ system at $M=m_{0}$.
The resolution is about 4(8)$\ \mevcc$ for $\Dzst$($\Dsm$)-tag and is asymmetric due to ISR effects. The resolution in $\Dzst$-tag method is less affected by ISR effects than the $\Dsm$-tag method. More events that emit ISR photons are removed in the $\Dzst$-tag method by the tighter signal window of $RM(K^+\Dzst)$ defined in Fig~\ref{fig:RM KD selection}. 
Thus, the resolution in the $\Dzst$-tag method becomes better than that in $\Dsm$-tag method.
The parameterization of the combinatorial-background shape is derived from the kernel estimate~\cite{Cranmer:2000du} of the mixed-background samples, which have been normalized to the estimated numbers as listed in Table~\ref{tab:bkg}.
The components of the NR signals and $\DDhs$ are included with shapes derived from MC simulations.
The sizes of the NR signals are free parameters in the fit, while the contributions from $\DDhs$ are fixed to the numbers in Table~\ref{tab:bkg}.

\begin{table*}[htbp]
    \caption{The fitted $\zcspm$ yields ($N_{\rm obs}$), the detection efficiencies ($\hat{\varepsilon}$), the integrated luminosities ($\mathcal{L}^{\rm int}$), the vacuum polarization correction factors ($1+\delta_{\rm VP}$) and the results for $\sigma^{\rm obs}\cdot\mathcal{B}$ for the different energy points. The uncertainties are statistical only.}
    \begin{center}\begin{tabular}{cccccccc}
    \hline\hline
    $\sqrt{s}\ (\gev)$ &Tag &$N_{\rm obs}$ &$\hat{\varepsilon}\ (\%)$ &$\mathcal{L}^{\rm int}\ ({\rm pb}^{-1})$ &$(1+\delta_{\rm VP})$ &$\sigma^{\rm obs}\cdot\mathcal{B}\ ({\rm pb})$ \\
    \hline
    \multirow{2}*{4.661} &$\Dsm$-tag  & $0^{+1.7}_{-0}$ &$0.1578\pm0.0009$ &\multirow{2}*{529.63}   &\multirow{2}*{1.055} &\multirow{2}*{$0.0^{+0.3}_{-0.0}$} \\
                        &$\Dzst$-tag & $0^{+1.3}_{-0}$ &$0.0293\pm0.0004$ \\
    \hline
    \multirow{2}*{4.682} &$\Dsm$-tag  & $25^{+7}_{-7}$  &$0.2017\pm0.0009$ &\multirow{2}*{1669.31}  &\multirow{2}*{1.055} &\multirow{2}*{$1.2^{+0.3}_{-0.3}$} \\
                        &$\Dzst$-tag & $18^{+5}_{-5}$ &$0.0363\pm0.0004$ \\
    \hline
    \multirow{2}*{4.699} &$\Dsm$-tag  & $7^{+5}_{-4}$  &$0.2151\pm0.0009$ &\multirow{2}*{536.45}   &\multirow{2}*{1.055} &\multirow{2}*{$0.9^{+0.6}_{-0.5}$} \\
                        &$\Dzst$-tag & $4.6^{+3.1}_{-2.6}$ &$0.0347\pm0.0004$ \\
    \hline\hline
    \end{tabular}\end{center}
    \label{tab:fit}
\end{table*}

\begin{table*}[!htbp]
    \caption{The upper limits for $\sigma^{\rm Born}_{\rm upper}\cdot\mathcal{B}$ in unit of pb at the 90\% confidence level at $\sqrt{s}=4.661$, 4.682, and $4.699\ \gev$. Here, both statistical and systematic uncertainties are taken into account.}
    \begin{center}\begin{tabular}{cccccc}
    \hline\hline
    \begin{tabular}[c]{@{}c@{}}$m_0(\zcsprime)$\\$(\mevcc)$\end{tabular} &$\Gamma_0=10\ \mev$ &$\Gamma_0=20\ \mev$ &$\Gamma_0=30\ \mev$ &$\Gamma_0=40\ \mev$ &$\Gamma_0=50\ \mev$  \\
    \hline\hline
    \multicolumn{6}{c}{at $\sqrt{s}=4.661\ \gev$}\\
    4120 &1.8 &1.9 &1.9 &2.0 &2.0  \\
    4125 &2.8 &2.6 &2.6 &3.4 &2.4  \\
    4130 &4.0 &3.8 &3.5 &3.2 &2.9  \\
    4135 &3.7 &4.0 &4.1 &5.6 &4.0  \\
    4140 &4.4 &4.0 &4.2 &4.3 &4.3  \\
    \hline
    \multicolumn{6}{c}{at $\sqrt{s}=4.682\ \gev$}\\
    4120 &3.3 &3.6 &3.7 &3.8 &3.8  \\
    4125 &4.5 &4.7 &4.9 &4.9 &4.8  \\
    4130 &4.3 &5.0 &5.2 &5.2 &5.2  \\
    4135 &3.9 &5.5 &6.2 &6.3 &6.3  \\
    4140 &2.8 &4.5 &5.8 &6.4 &6.6  \\
    \hline
    \multicolumn{6}{c}{at $\sqrt{s}=4.699\ \gev$}\\

    4120 &4.1 &4.8 &4.9 &5.0 &5.0  \\
    4125 &5.1 &5.3 &5.6 &5.7 &5.8  \\
    4130 &4.1 &4.5 &4.9 &5.2 &5.2  \\
    4135 &4.4 &5.2 &5.7 &6.1 &6.1  \\
    4140 &2.7 &3.8 &4.6 &5.4 &5.4  \\
    \hline\hline
    \end{tabular}\end{center}
    \label{tab:xsupper}
\end{table*}

\begin{figure*}[htbp!]
\centering
\includegraphics[width=0.31\textwidth]{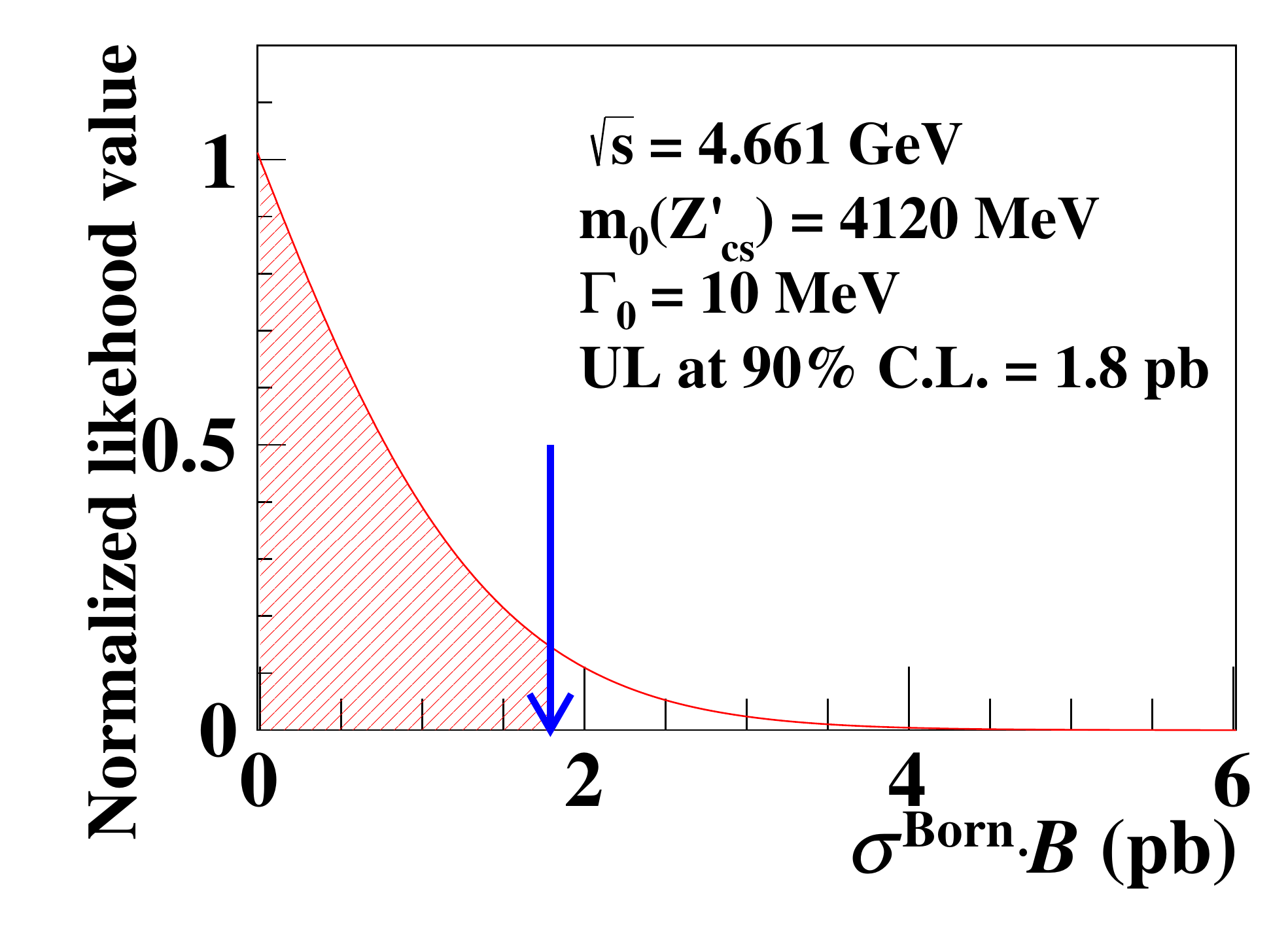}
\includegraphics[width=0.31\textwidth]{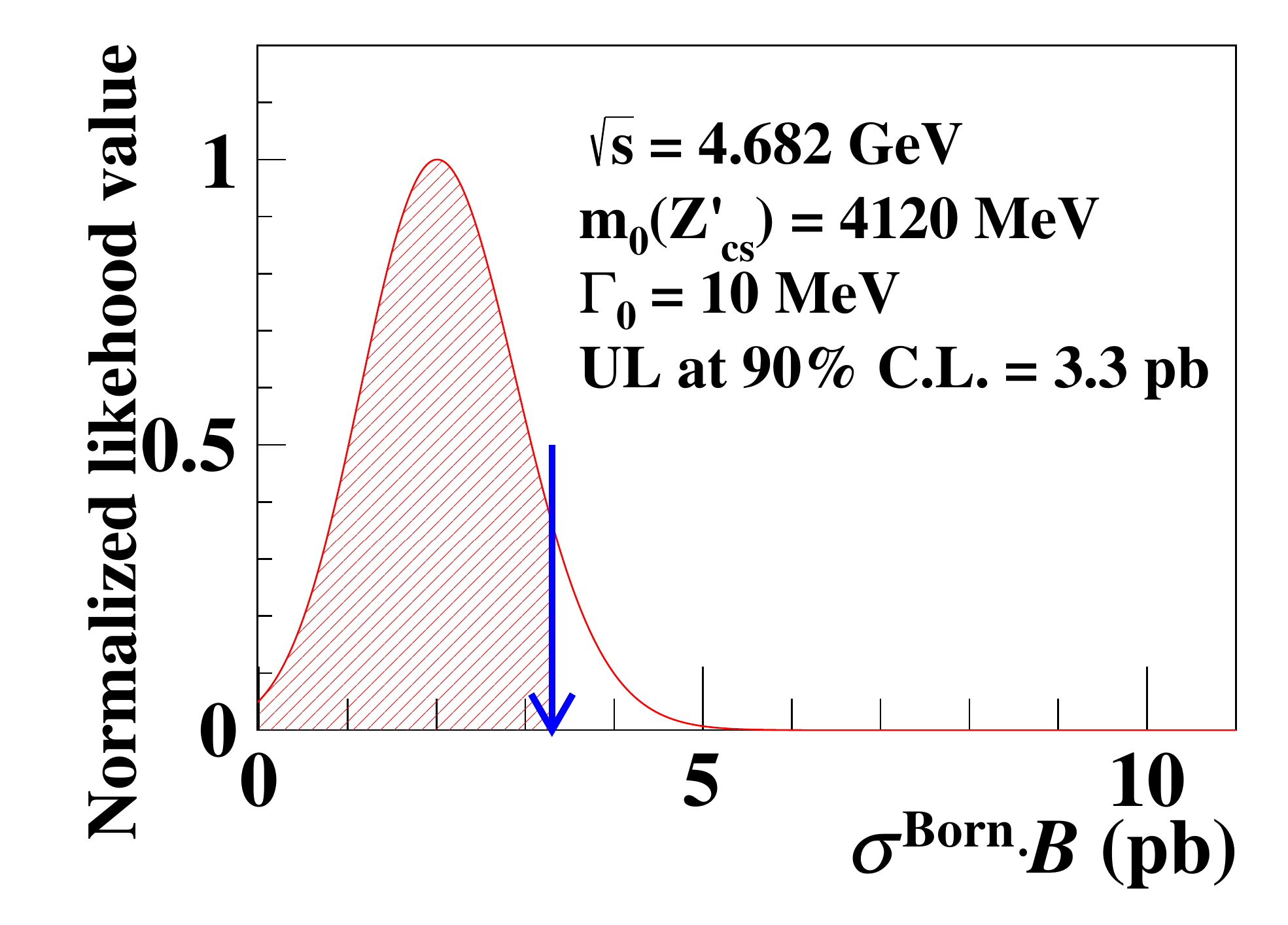}
\includegraphics[width=0.31\textwidth]{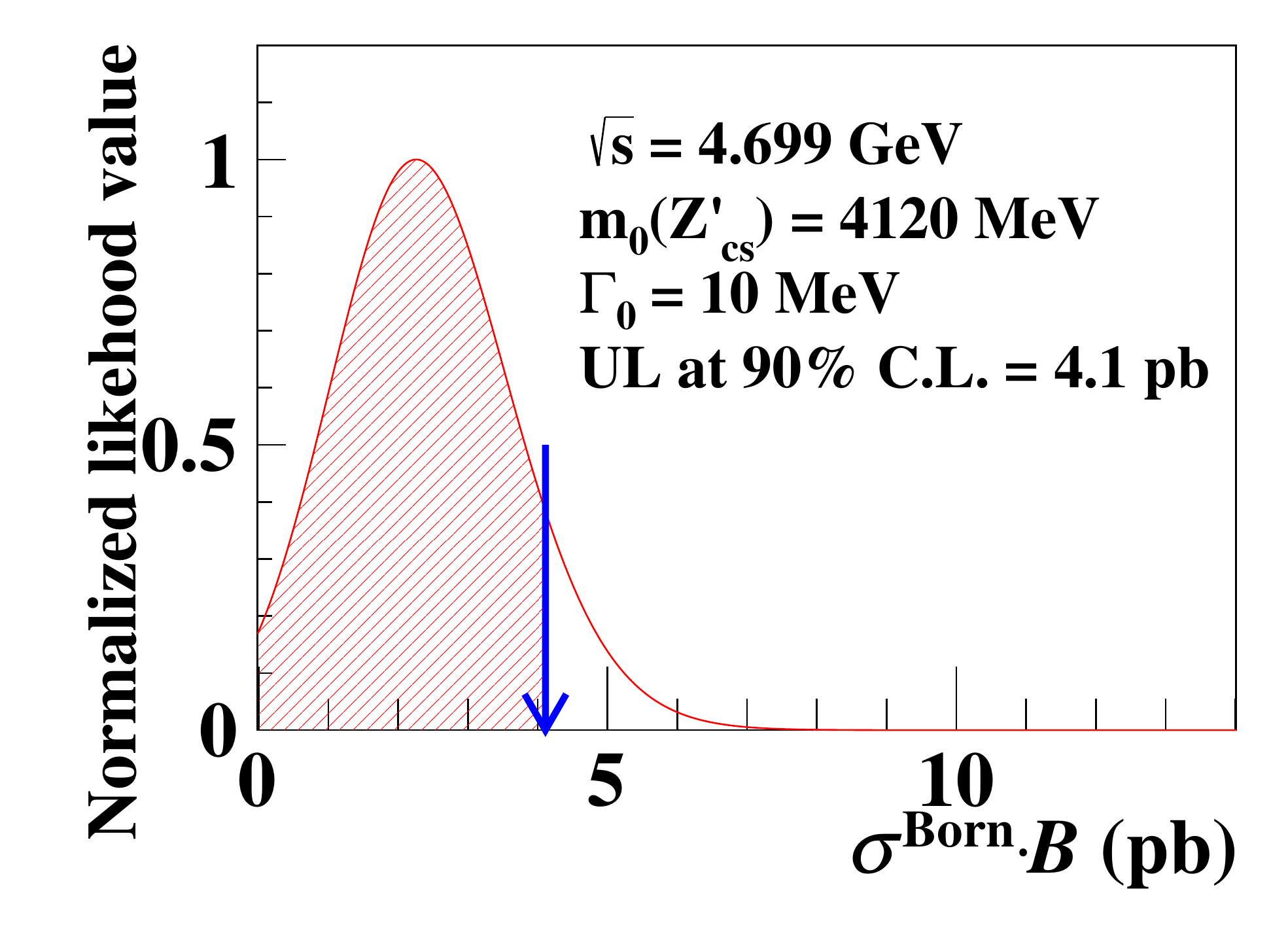}
\caption{
Likelihood scan of $\sigma^\mathrm{Born}\cdot\mathcal{B}$ for the data samples at $\sqrt{s}=4.661, 4,682$, and $4.699\ \gev$ after considering systematic uncertainties.
The blue arrow indicates the upper limit at the 90\% confidence level. In the likelihood scan, the PDFs which result in the largest upper limits are used, and the mass and width of $Z_{cs}'$ are fixed to $4120\ \mevcc$ and $10\ \mev$, respectively.}
\label{fig:scan UL}
\end{figure*}

\begin{figure*}[tph]
\centering
\includegraphics[width=1.0\textwidth]{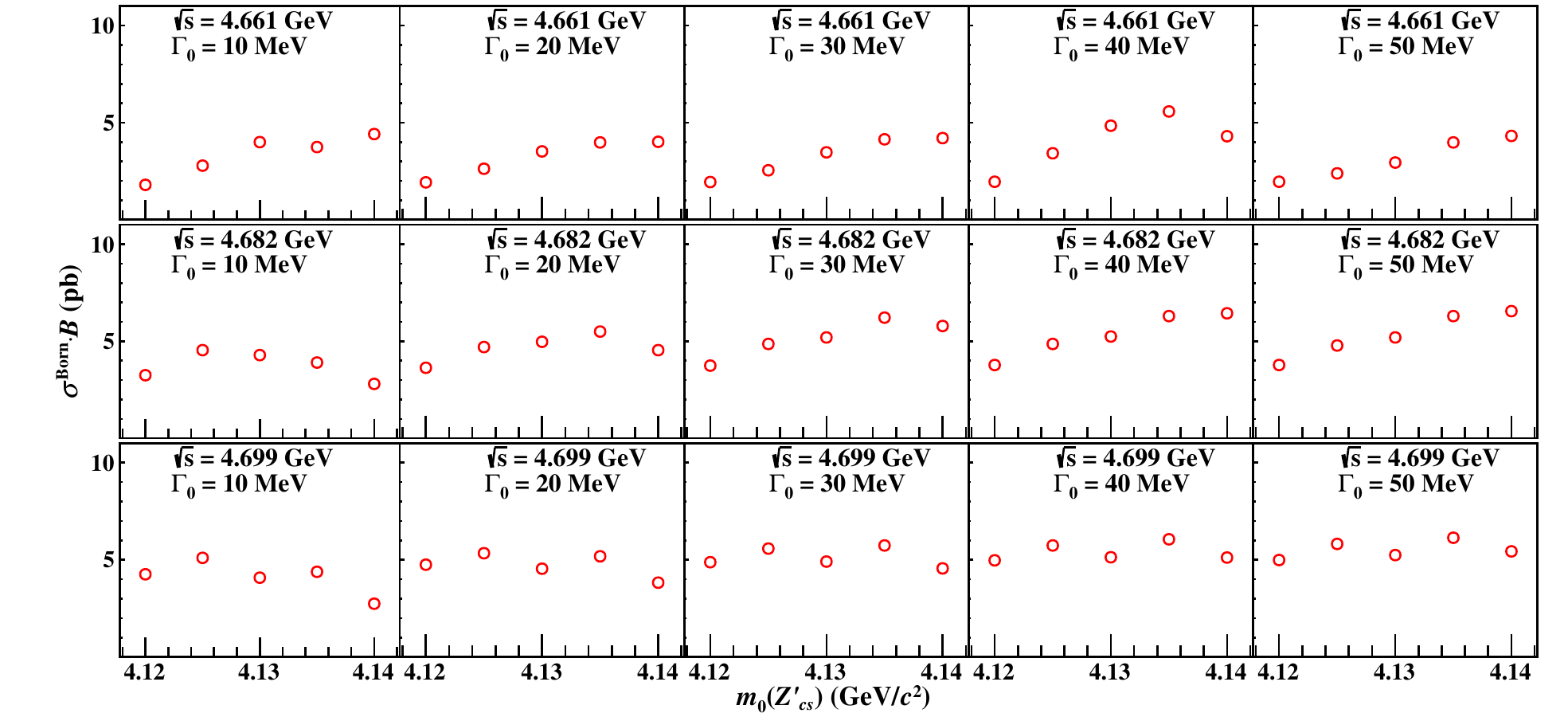}
\caption{
Upper limits (open red circles) for $\sigma^{\rm Born}_{\rm upper}\cdot\mathcal{B}$ at the 90\% confidence level for the assumed masses of $m_0(\zcsprime)=4120, 4125, 4130, 4135$ and 4140 MeV$/c^{2}$, and the assumed widths of $\Gamma_0=10, 20, 30, 40$ and $50 \mev$
at $\sqrt{s}=4.661$, 4.682 and $4.699\ \gev$.
}
\label{fig:xsupper}
\end{figure*}

From the simultaneous fit, the mass of $\zcspm$ is determined to be $m_{0}=(4123.5\pm0.7)\mevcc$.
The statistical significance of the signal is $3.9\sigma$, which is calculated with 5 degrees of freedom and $-2\ln(L_0/L_\mathrm{max})=25.8$, where $L_0$ and $L_\mathrm{max}$ are the fitted likelihood values with and without involving the $Z_{cs}^{\prime-}$ signal component in the fit, respectively. The degrees of freedom consist of the signal yields at the three center-of-mass energies and the mass and width of the $Z_{cs}'$ state. After taking into account the systematic uncertainties, the significance is reduced to $2.1\sigma$. 
The width $\Gamma_{0}$ is not reported as the size of the data samples is inadequate to determine this parameter in the fit. 
Alternatively, we perform a series of local $p$-value scans under different width assumptions $(10, 20, 30, 40, 50)\mev$ in the $RM(\kp)$ spectra, as shown in Fig.~\ref{fig:pscan}.
The minimum $p$-value is found to be $4.0\times10^{-5}$ at $m_{0}=4124.1\mevcc$ with $\Gamma_{0}=10\mev$, corresponding to a local statistical significance of 4.1\,$\sigma$.

\begin{center}
\includegraphics[width=0.45\textwidth]{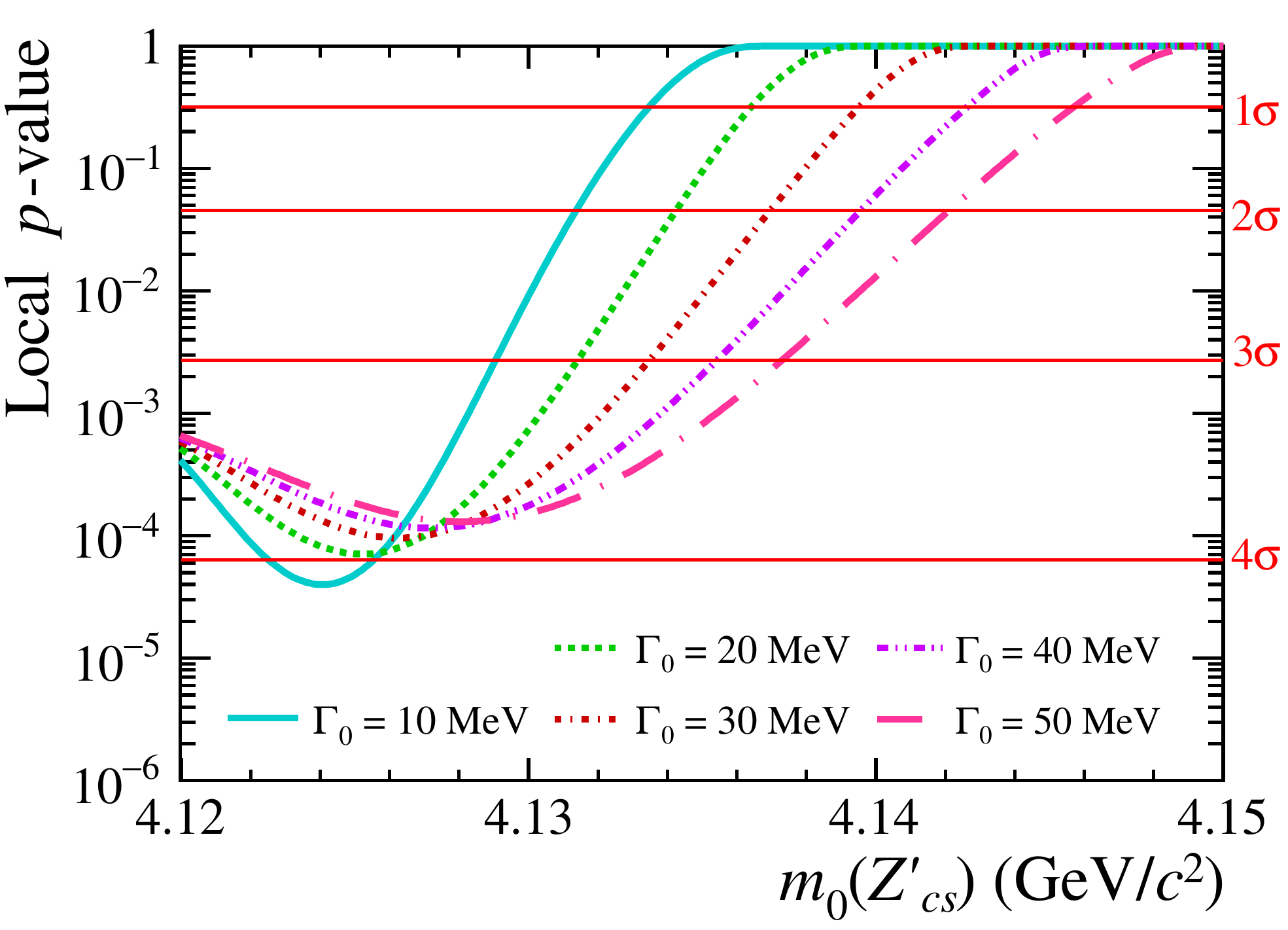}
\figcaption{The local $p$-value as a function of $m_0(\zcsprime)$ from the simultaneous fit to data samples at $\sqrt{s}=4.661, 4.682$, and $4.699\ \gev$.}
\label{fig:pscan}
\end{center}
\section{Measurement of Born cross sections}
The product of the Born cross section and the branching fraction of $\zcspm\rightarrow \Dsstm\Dzst$, $\sigma^{\rm Born}(\ee\rightarrow\kp\zcspm+c.c.)\cdot\mathcal{B}(\zcspm\rightarrow\Dsstm\Dzst)$($\sigma^{\rm Born}\cdot\mathcal{B}$ for short), is determined by
\begin{eqnarray}
    \sigma^{\rm Born}\cdot\mathcal{B}=\frac{N_{{\rm obs}, i}^{k}}{\mathcal{L}_{i}^{\rm int}\cdot\hat{\varepsilon}_{i}^{k}\cdot(1+\delta_{\rm ISR})_{i}\cdot(1+\delta_{\rm VP})_{i}}.
    \label{eq:xs}
\end{eqnarray}
Here $N_{{\rm obs}, i}^{k}$ is the $\zcspm$ signal yield from the fit shown in Fig.~\ref{fig:RM_K fit};
$\hat{\varepsilon}_{i}^{k}$ is the combined branching-fraction-weighted detection efficiency in different $D_{s}^{-}(\Dzst)$ decay modes;
$\mathcal{L}_{i}^{\rm int}$ is the integrated luminosity;
$(1+\delta_{\rm ISR})_{i}$ is the ISR correction factor;
$(1+\delta_{\rm VP})_{i}$ is the vacuum-polarization correction factor, which is calculated according to Ref.~\cite{Jegerlehner:2017zsb};
and $i$ and $k$ denote the energy point ($1\le i \le 3$) and tag method ($1\le k \le 2$), respectively.
The measured values of $\sigma^{\rm obs}\cdot\mathcal{B}$ are listed in Table~\ref{tab:fit}.

We furthermore determine the upper limits of $\sigma^{\rm Born}\cdot\mathcal{B}$ at the 90\% confidence level at all three energy points for different values of the $\zcspm$ mass and width by measuring the signal yields under these different assumptions.
The upper limits are obtained via a likelihood scan, as illustrated in Fig.~\ref{fig:scan UL}.
The systematic uncertainties are taken into account in the upper-limit estimation by adopting a Gaussian constraint method~\cite{gaussian} in the likelihood scan. 
Here, to be conservative, the ISR correction factor $(1+\delta_{\rm ISR})_{i}$ are evaluated based on a hypothetical narrow resonant line shape of the cross sections with a width of $10\mev$ following the method in Ref.~\cite{BESIII:2020tgt}, which gives a minimum ISR correction factor. The ISR correction factors as a function of $\sqrt{s}$ is shown in Fig.~\ref{fig:kappa}.
The upper limits of $\sigma^{\rm Born}\cdot\mathcal{B}$ after accounting for systematic uncertainties are shown in Fig.~\ref{fig:xsupper} and Table~\ref{tab:xsupper}.

\section{Systematic uncertainty}
The systematic uncertainties on the mass measurement of the $\zcspm$ and the upper-limit cross sections are summarized in Table~\ref{tab:sys_mass} and Table~\ref{tab:sys_xs}, respectively.
For the mass measurement, the systematic uncertainties are mainly from the estimation of combinatorial background and the $\DDhs$ contribution, the input Born cross-section line shape for $\sigma^{\rm Born}(\ee\rightarrow\kp\zcspm)$, detector resolution, mass scaling and signal model. 
These sources of uncertainty are augmented by additional contributions in the cross-section measurement.

\subsection{Systematic uncertainties on the mass measurement}
\begin{center}
   \tabcaption{Absolute systematic uncertainties in the measurement of the $\zcspm$ mass. }
   \begin{tabular}{lc}
	\hline \hline
	Source & Mass $(\mathrm{MeV}/c^{2})$ \\
	\hline
	Comb. background                & $0.1$    \\ 
    $\DDhs$                         & $0.1$    \\
    $\sigma^{\rm Born}(\ee\to K^+\zcspm)$ line shape                 & $0.5$    \\
    Signal model                    & $0.1$    \\
    Mass scaling                    & $0.5$    \\
    Resolution                      & $0.8$    \\
    Efficiency curve                & $<0.1$\hspace*{0.3cm} \\ 
    $\Gamma_0$ assumptions          & $4.6$ \\
    \hline
    Total                           & $4.7$    \\
  	\hline \hline
	\label{tab:sys_mass} 
\end{tabular}
\end{center}

The systematic uncertainties related to the combinatorial backgrounds and $\DDhs$ contributions are estimated by floating their sizes with a Gaussian constraint in the fit.
The mean and width of the Gaussian function are derived according to the estimated yields and the corresponding uncertainties, as listed in Table~\ref{tab:bkg}.
Both show variation of $0.1\mevcc$ on the fitted mass, which is taken as the systematic uncertainty.

We replace the input Born cross-section line shape of $\sigma^{\rm Born}(\ee\rightarrow\kp\zcspm)$ in MC production with a narrow Breit-Wigner line-shape assumption and repeat the signal determination. The width of the narrow Breit-Wigner line-shape is chosen to be $10\mev$, which is much narrower than the nominal. The change of $0.5\mevcc$ for the fitted mass is assigned as the systematic uncertainty.

In the nominal fit, the signal model is based on the spin-parity assumption of $\zcspm$ as $J^{P}=1^{+}$ and the assumption that the relative momentum between $\kp$ and $\zcspm$ in the rest frame of the $\ee$ system and the relative momentum between $\Dsstm$ and $\Dzst$ in the $\zcspm$ system are both in an $S$-wave state, denoted as $1^{+}(S, S)$. These assumptions are consistent with the study of $\zcsm$~\cite{BESIII:Zcs}.
As systematic variations, we examine the assumptions of spin-parity and angular momentum with $0^{-}(P,P), 1^{-}(P, P), 1^{+}(D, S), 2^{-}(P, P)$ and $2^{+}(D, S)$ configurations.
The maximum change of $0.1\mevcc$ for the fitted mass is assigned as the systematic uncertainty for the spin-parity assumption.

We select a control sample of $\ee\rightarrow\DDhs$, as done in Ref.~\cite{BESIII:Zcs}, to study the mass scaling of the recoil mass of the low-momentum bachelor $\kp$.
An MC-determined signal shape convolved with a Gaussian function is used to represent the difference between data and MC simulation.
The mean and width of this function are determined to be $-0.2\pm0.5\mevcc$ and $<1.43\mevcc$ (68\% confidence level), respectively.
We take the maximum mass shift of $\sqrt{0.2^{2}+0.5^{2}}\approx0.5\mevcc$ as the systematic uncertainty due to the mass scaling.

To estimate systematic uncertainties associated with resolution smearing function, which reflects imperfect MC simulation for the resolution, we smear the resolution function in the fit to the $\zcspm$ spectrum and re-perform the mass fit. The mean and width of the Gaussian smearing function are also determined from the control sample of $\ee\rightarrow\DDhs$, which are $0.5\mevcc$ and $1.43\mevcc$.
The resultant change in the fitted mass, $0.8\mevcc$, is taken as the systematic uncertainty.

The change in the fitted mass is found to be negligible when the efficiency curves adopted in the resonance fit are varied within the uncertainties of their parametrizations. 

As previously mentioned, we perform a series of scans under different width assumptions since the $\Gamma_0$ of the $Z_{cs}'$ state is not determined due to the limited size of the data samples.
Under different width assumptions, the fitted mass can be different. Thus, we assign a systematic uncertainty, $4.6\mevcc$, on the measured mass according to the largest discrepancy between the nominal fit result and the fitted mass with the width fixed to the assumed widths.

\subsection{Systematic uncertainties on the upper-limit cross sections}
\begin{table*}[htbp]
    \caption{Relative systematic uncertainties in the measurement of $\sigma^{\rm Born}\cdot\mathcal{B}$. The horizontal dividing lines separate the three categories that are described in the main text.}
    \begin{center}\begin{tabular}{ccccccc}
        \hline\hline
        \multirow{2}{*}{Source}                       &\multicolumn{3}{c}{$\Dsm$-tag } &\multicolumn{3}{c}{$\Dzst$-tag}\\
        \cmidrule(lr){2-4} \cmidrule(lr){5-7}
                                                      &4.661 GeV &4.682 GeV &4.699 GeV &4.661 GeV &4.682 GeV &4.699 GeV\\
        \hline
        Tracking                          &$3.6\%$ &$3.6\%$ &$3.6\%$ &$4.1\%$ &$4.1\%$ &$4.1\%$ \\
        PID	                              &$3.6\%$ &$3.6\%$ &$3.6\%$ &$4.1\%$ &$4.1\%$ &$4.1\%$ \\
        $K_{S}^0$($\pi^0$) reconstruction &$0.4\%$ &$0.4\%$ &$0.4\%$ &$2.0\%$ &$2.0\%$ &$2.0\%$ \\
        $RM(K^+\Dsm(\Dzst))$                &$0.2\%$ &$0.2\%$ &$0.2\%$ &$6.1\%$ &$5.1\%$ &$4.3\%$ \\
        Luminosity                       &$1.0\%$ &$1.0\%$ &$1.0\%$ &$1.0\%$ &$1.0\%$ &$1.0\%$ \\
        Input branching fractions                         &$2.3\%$ &$2.3\%$ &$2.3\%$ &$1.0\%$ &$1.0\%$ &$1.0\%$ \\
        \hline
        Comb. bkg                &\multicolumn{6}{c}{\multirow{2}{*}{See Table~\ref{tab:bkg} for the background components.}}\\
        $\DDhs{}$                            \\
        \hline
        Efficiency curve                             &\multicolumn{6}{c}{\multirow{4}{*}{See main text for discussion of the treatment of these four systematic effects.}} \\
        Line shape                                    \\
        Resolution                                    \\
        Signal model                                  \\
        \hline\hline
    \end{tabular}\end{center}
    \label{tab:sys_xs}
\end{table*}
We consider three categories of systematic uncertainties on the upper limit of $\sigma^{\rm Born}_{\rm upper}\cdot\mathcal{B}$.

Category I includes items such as the detection efficiency, the integrated luminosities and the input branching fractions, whose uncertainties affect the cross-section measurement multiplicatively.
Average uncertainties associated with the tracking, PID and $K_{S}^{0}(\pi^{0})$ reconstruction efficiencies are estimated to be 3.6\%, 3.6\% and 0.4\% (4.1\%, 4.1\% and 2.0\%) in $\Dsm(\Dzst)$-tag method, respectively. 
The efficiency of the $RM(\kp\Dsm)$ and $RM(\kp\Dzst)$ requirement is re-estimated by changing the MC-simulated resolution according to the observed difference with respect to data.
The resulting changes, 0.2\% in the $\Dsm$-tag method, and 6.1\%, 5.1\% and 4.3\% in the $\Dzst$-tag method, at various energy points are taken as the systematic uncertainties. The differences in systematic uncertainties here are due to that the requirement of $RM(\kp\Dzst)$ is much tighter than $RM(\kp\Dzst)$, which makes the $\Dzst$ method more sensitive to the resolution changes. 
The integrated-luminosity uncertainty, measured with large-angle Bhabha scattering events, is estimated to be 1\%.
The uncertainties on the cited branching fractions for the processes in the reconstructed decay chains are evaluated~\cite{ParticleDataGroup:2020ssz}.
The total multiplicative uncertainty is obtained by adding the uncertainties of all these components in quadrature. In the estimation of the upper limit, the multiplicative uncertainty is taken into account through one single Gaussian constraint on the product of the central value of the multiplicative components.

Category II comprises the combinatorial background and $\DDhs$ uncertainties, whose shapes and sizes are fixed in the nominal fit.
The uncertainties on their sizes, as listed in Table~\ref{tab:bkg}, affect the estimation of the upper limit of $\sigma^{\mathrm{Bron}}_{\mathrm{upper}}\cdot\mathcal{B}$. 
To take into account their effects, the sizes of these contributions are floated in the upper-limit estimation with Gaussian constraints on their uncertainties~\cite{gaussian}. 
In addition, to consider the effects from the background-shape parametrization, a bootstrap re-sampling method is carried out to generate 500 background toy samples.
The background template shape from each toy sample is adopted in each new fit and the distribution of the corresponding differences from the nominal fit is parameterized by a Gaussian function.
The likelihood-scan curves are smeared with this Gaussian function to include the systematic effect.

 \begin{center}
    \includegraphics[width=0.35\textwidth]{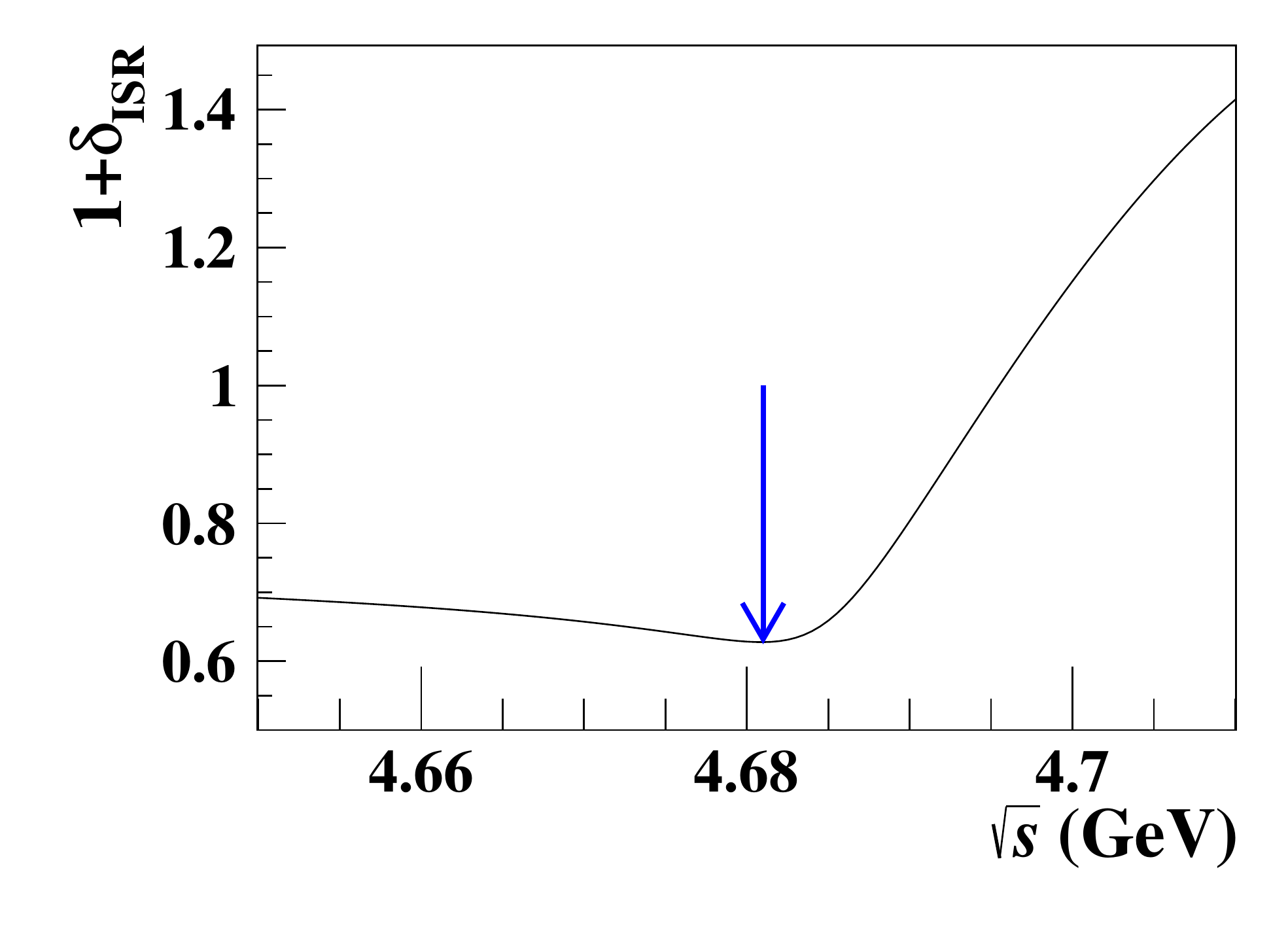}
    \figcaption{The corresponding ISR correction factor $(1+\delta_{\mathrm{ISR}})$ as a function of $\sqrt{s}$ for a narrow resonant line shape. We take the minimum of $(1+\delta_\mathrm{ISR})$, as indicated by the blue arrow, to obtain a conservative estimation of $\sigma^{\rm Born}_{\rm upper}\cdot\mathcal{B}$.}
    \label{fig:kappa}
\end{center}

Category III comprises other sources of uncertainty, such as the efficiency parametrization, the input line shape of the cross sections, the detector resolution and the signal model, for which we evaluate the size of possible bias directly by repeating the fit under different assumptions.
The effect from the efficiency parametrization is estimated by varying the parametrization within its fit uncertainties, and the differences in cross sections at each energy point are taken as the corresponding systematic uncertainty.
Following Ref.~\cite{BESIII:2020tgt}, we consider the influence from the input line shape on the estimation of the ISR correction factors by choosing a minimum value of $(1+\delta_{\rm ISR})$, as shown in Fig.~\ref{fig:kappa}, under a narrow resonant line shape assumption. 
This provides a conservative estimation on the upper limit.
For the resolution effect, we increase the width of the smearing function by $1.43 \mev$, which is estimated based on the control sample of $\ee\rightarrow\DDhs$ events as previously mentioned, to accommodate potential differences in the detector resolution between data and MC simulation.
For the signal model, we obtain various upper limits of $\sigma^{\rm Born}_{\rm upper}\cdot\mathcal{B}$ based on six different spin-parity assignments, as considered for the mass measurement.
For all the above variations, the maximum upper limits, after considering the systematic uncertainties of categories I and II, are assigned as the final results for the different resonance parameter hypotheses at the various energy points.
The upper limits of $\sigma^{\rm Born}\cdot\mathcal{B}$, after taking account of the systematic uncertainties, are shown in Fig.~\ref{fig:xsupper} and Table~\ref{tab:xsupper}.

\subsection{Systematic uncertainties on the signal significance}
The systematic uncertainties on the determination of the signal significance arise from the estimation of combinatorial background and $\DDhs$ contribution, the choice of signal model and the width assumptions.
In each signal model aforementioned, we change the fixed sizes of combinatorial background and $\DDhs$ contribution by $\pm 1 \sigma$ and then calculate the significance by performing the fits with and without signal components. We adopt 2.1$\sigma$, which is the smallest significance observed from this procedure, to constitute the result including systematic uncertainties.

\section{Summary}
Motivated by the observation of $\zcs^-$, we search for its highly excited partner $\zcspm$ based on the $\ee$ annihilation data collected at $\sqrt{s}=4.661, 4.682$ and $4.699\ \gev$ with the BESIII detector. By using a partial-reconstruction technique, the $K^+$ recoil mass spectra are used to search for the $\zcspm$ state in $\DDstar$ events.
We find a small excess of $\zcspm\to\DDstar$ events with a significance of $2.1\sigma$, after considering systematic uncertainties, and measure its mass to be $m_0(\zcsprime) = (4123.5\pm0.7_\mathrm{stat.}\pm4.7_\mathrm{syst.})\mevcc$
from a simultaneous fit to all data samples. 
However, our data set is too small in size to make a meaningful measurement of the width of this state.
Hence, a series of $\zcspm$ width hypotheses are tested and the corresponding $p$-values are evaluated, as shown in Fig.~\ref{fig:pscan}. 
A local minimum $p$-value for $\zcspm$ is found at $m_{0}=4124.1\mevcc$ when setting $\Gamma_{0}=10$ MeV, corresponding to a local statistical significance of 4.1$\sigma$.

Due to the limited sample size, we report the upper limits at the 90\% confidence level of the product of the Born cross sections and the branching fraction of the decays $\sigma^{\rm Born}(\ee\rightarrow\kp\zcspm+c.c.)\cdot\mathcal{B}(\zcspm\rightarrow\Dsstm\Dzst)$ at $\sqrt{s}=4.661$, $4.682$, and $4.699\ \gev$. The search is performed with the mass of $\zcspm$ assumed to be $4.120$, $4.125$, $4.130$, $4.135$ and $4.140\ \gevcc$, and the width assumed to be $10$, $20$, $30$, $40$ and $50\ \mev$. The upper limits are $2\sim6$, $3\sim7$ and $3\sim6$ pb for $\sqrt{s}=4.661$, $4.682$, and $4.699\ \gev$, respectively, as shown in Fig.~\ref{fig:xsupper} and Table~\ref{tab:xsupper}. The limited statistical precision prevents the establishment of the potential $\zcspm$ state, but the further data taking foreseen in the region $\sqrt{s}=4.6\sim4.9\ \gev$~\cite{BESIII:2020nme} may allow for a clearer picture to emerge.

\acknowledgments{The BESIII collaboration thanks the staff of BEPCII and the IHEP computing center for their strong support. }

\end{multicols}

\vspace{-1mm}
\centerline{\rule{80mm}{0.1pt}}
\vspace{2mm}

\begin{multicols}{2}
%

\end{multicols}

\clearpage
\end{CJK*}
\end{document}